\documentclass{IEEEtran}
\usepackage[utf8]{inputenc}
\usepackage{amsmath}
\usepackage{graphicx}
\usepackage{hyperref}
\usepackage[noadjust]{cite}

\title{Spatially-Aware Temporal Anomaly Mapping of Gamma Spectra}
\author{Alex~Reinhart, Alex~Athey, and Steven~Biegalski%
  \thanks{Manuscript received April 3, 2013; revised August 21, 2013; accepted
    April 9, 2014. This research has been funded in part by the Office of the
    Assistant Secretary of Defense for Nuclear Matters.}%
  \thanks{A.~Reinhart is with the Department of Statistics, Carnegie Mellon
    University, Pittsburgh PA 15213 (e-mail areinhar@stat.cmu.edu).}%
  \thanks{A.~Athey is with Applied Research Laboratories, The University of
    Texas at Austin, Austin TX 78713 (e-mail alex.athey@arlut.utexas.edu).}%
  \thanks{S.~Biegalski is with the Nuclear Engineering Teaching Laboratory, The
    University of Texas at Austin, TX 78712.}%
  \thanks{Accepted for publication in \textit{IEEE Transactions on Nuclear
      Science}, DOI 10.1109/TNS.2014.2317593. © 2014 IEEE. Personal use of this
    material is permitted. Permission from IEEE must be obtained for all other
    uses, in any current or future media, including reprinting/republishing this
    material for advertising or promotional purposes, creating new collective
    works, for resale or redistribution to servers or lists, or reuse of any
    copyrighted component of this work in other works.}}

\date{\today}

\begin{document}
\maketitle
\begin{abstract}
  For security, environmental, and regulatory purposes it is useful to
  continuously monitor wide areas for unexpected changes in radioactivity. We
  report on a temporal anomaly detection algorithm which uses mobile detectors
  to build a spatial map of background spectra, allowing sensitive detection of
  any anomalies through many days or months of monitoring. We adapt
  previously-developed anomaly detection methods, which compare spectral shape
  rather than count rate, to function with limited background data, allowing
  sensitive detection of small changes in spectral shape from day to day. To
  demonstrate this technique we collected daily observations over the period of
  six weeks on a 0.33 square mile research campus and performed source injection
  simulations.
\end{abstract}

\section{Introduction}

\IEEEPARstart{C}{onducting} wide-area radiation surveillance is a challenging
problem for environmental and security applications. For example, a city may
wish to produce a map of radiation sources over a wide region, identify
unexpected or unauthorized radioactive sources and take corrective action, then
monitor the area for any future introduction of sources. Dedicated systems have
been developed, such as the United States Department of Energy's Aerial
Measuring System, which uses aircraft to map radiological activity at nuclear
sites and during emergencies \cite{Wasiolek:2007vn,AMS}; however, these systems
focus on one-time mapping, rather than continuous monitoring and surveillance.

For city-sized areas, current mapping efforts typically use low-flying
helicopters.  These operations produce high-resolution maps, but require several
weeks of intensive flying, making them unsuitable for continuous
monitoring. Other methods of mapping include gamma-ray imaging devices mounted
in vehicles \cite{Zelakiewicz:2011ig}; mobile scintillator detectors which
search for sudden changes in background spectral shape, indicating the detector
is traveling past a source \cite{Pfund:2007,Pfund:2010hm}; and
commercially-available portable identifying spectrometers.

These methods are limited in a number of ways. An aerial survey which requires
weeks to complete cannot continuously monitor a wide area in real time. For
vehicle-based mobile detectors, previous work has developed spectroscopic source
detection algorithms, but these do not perform change detection from one survey
to the next, and do not take advantage of previous background observations to
improve detection sensitivity \cite{Pfund:2010hm}. Imaging methods require large
and expensive detectors and focus on detecting and imaging individual sources,
rather than mapping large areas.

We report on an integrated system of (a) mobile data collection, (b)
spatial-temporal database of gamma spectra, and (c) spatially-aware anomaly
mapping algorithm. Collectively we refer to this system as the spectral
comparison ratio anomaly mapping (SCRAM) system. The SCRAM system uses mobile
detectors to build spectral maps of wide areas and identifies temporal anomalies
using spectral comparison ratios (SCRs), detecting changes in spectral shape. By
using a spatial database of observed spectra and comparing new observations to
the recorded background, a high sensitivity to temporal anomalies is obtained.
To demonstrate these concepts we collect and analyze a sample dataset on a small
research campus.

This research is motivated as a feasibility study to consider collecting
radiation maps with vehicles that naturally transit areas of interest through
their daily operations, such as patrol vehicles, buses, or unmanned aerial
vehicles. Such a system, collecting data continuously as vehicles operate, would
provide high-sensitivity, continuous, wide-area surveillance at low operational
cost.

\section{SCRAM System}
The SCRAM system is composed of three core components: a mobile detection
platform, based on a scintillator detector and GPS system; a spatial and
temporal database of observed spectra allowing analysis and change detection;
and an analysis method based on spectral comparison ratios. We constructed a
proof-of-concept platform using off-the-shelf components.

\subsection{Mobile Detection Platform} 
The first component in the SCRAM system is a platform for collecting radiation
data. For data collection, the required electronics were USB-based GPS,
USB-based scintillator detector and a PC laptop; the largest of these components
was the laptop and all items were easily contained in a small backpack. This
could be carried by a person or placed in a golf cart or other vehicle. The
detector was a Bridgeport Instruments \(2\times2\)-inch cesium iodide
scintillator and was used to collect spectra continuously at two-second
intervals; however, the analysis system is agnostic to choice of sensor, and
would work with other gamma spectra detectors, including much larger or more
sensitive devices.

\subsection{Spatial-Spectral-Temporal Database}
The second component in the SCRAM system is a method of storing multiple-pass
radiation data in a spatial and temporal database.  For this, we constructed a
database of spectra using the PostGIS geographic information system
\cite{postgis,pgsql}. Software was developed to allow for rapid spatial and
temporal querying of spectra.  Spectra were recorded into the database along
with the detector's GPS coordinates, current time, and several diagnostic
fields.

We were able to use our spatial and temporal database to determine the expected
background at each location, along with the background variability with respect
to time. More importantly, new observations could be compared against the
background at a location by querying the database for past observations, which
were summed together to produce an accurate background spectrum determination.

\subsection{Spectral Comparison Ratio Anomaly Mapping}

The final component of the SCRAM system is a spectral anomaly detection
algorithm. Our algorithmic approach is based on the method of spectral
comparison ratios (SCRs), as previously developed by other authors
\cite{Pfund:2007,Jarman:2008fm}, with modification to accurately produce a
covariance estimate with limited data, such as the case where few observations
are made within a given spatial area.

The first step is to group data into square spatial cells. (We will refer to
components of the spectral histogram as ``bins'' and the spatial grid as
``cells'' to prevent confusion.) Data within each cell is summed, and the
newly-observed spectrum compared to the cell's recorded background
spectrum. Spatial cell size choices are discussed in Section~\ref{analysis}.

\subsubsection{Spectral Comparison Ratios}

\begin{figure}
  \centering
  \includegraphics[width=3.5in]{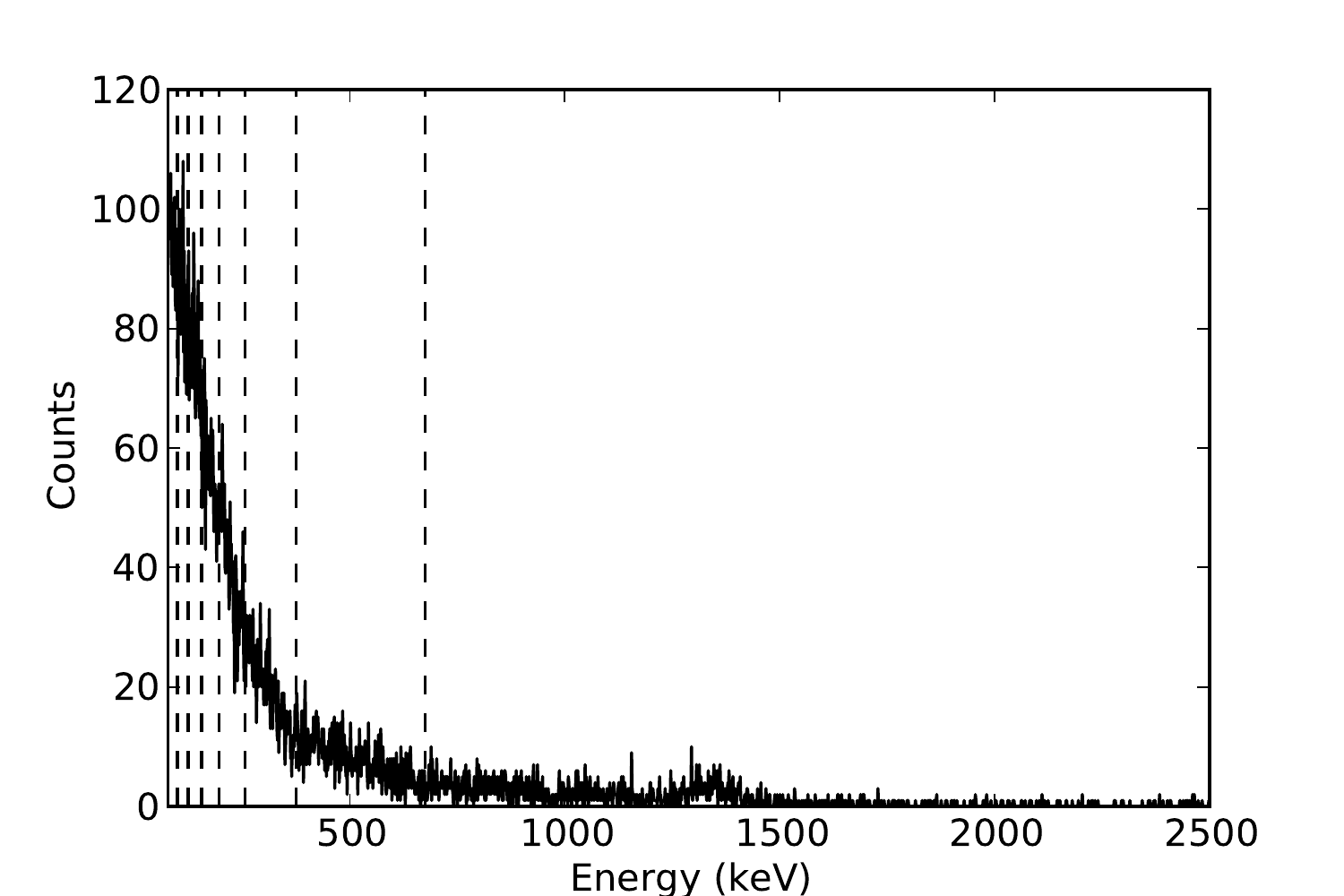}
  \caption{An example background spectrum, with vertical dashed lines indicating
    the boundaries of spectral bins. There are eight bins total.}
  \label{spectral-bins}
\end{figure}

The SCR method does not detect changes in radiation levels, but instead examines
spectral shapes. SCR collects observed gamma counts into energy bins, which may
be chosen to cover certain spectral regions of interest or can be distributed
evenly across the spectrum. Various methods exist to choose energy bins targeted
for detection of particular isotopes \cite{Wei:2010go,Pfund:2010hm}; however,
for this demonstration, we simply chose 8 energy bins which contained roughly
equal numbers of counts in a typical background spectrum, covering the energy
range between 100 and 2500 keV, as shown in
Fig.~\ref{spectral-bins}. Applications which are targeted to specific isotopes
or which wish to reject certain nuisance isotopes may use bins selected for this
purpose. In the course of this research, we examined a number of bin choices,
including bins with widths proportional to the square root of energy, as used in
previous research \cite{Jarman:2008fm}; the results of subsequent analysis were
qualitatively the same regardless of bin choice, although quantitively the
sensitivity to specific isotope detection varied with bin choice, as expected.

Binning the data into \(n\) energy bins creates a vector of counts \(\mathbf{c}
= [ c_1 c_2 \ldots c_n]\). We also have a background vector, \(\mathbf{b}\),
containing the observed background spectrum, binned in the same way. One energy
bin is chosen as a reference bin (in this case, bin 1). The choice of reference
bin is unimportant, as the anomaly statistic has been shown to be invariant to
the choice of bin \cite{Runkle:2009ev}. The spectral comparison ratios can then
be computed by
\begin{equation}\label{scr}
s_i = c_1 - \frac{b_1}{b_i} c_i,
\end{equation}
for \(i > 1\). Here \(c_1\) and \(b_1\) are the numbers of counts in the
reference bin. This is mathematically equivalent to multiplying the vector
\(\mathbf{c}\) by a spectral shape matrix \(S\):

\begin{equation}
S = \begin{pmatrix}
1 & - \frac{b_1}{b_2} & 0 & \ldots & 0 \\
1 & 0 & - \frac{b_1}{b_3} & \ldots & 0\\
\vdots \\
1 & 0 & 0 & \ldots & -\frac{b_1}{b_n} \\
\end{pmatrix}
\end{equation}

\begin{equation}
\mathbf{s} = S \cdot \mathbf{c}
\end{equation}

There are \((n - 1)\) linearly independent SCRs, since one bin is a reference
bin.  The SCR process compares \(c_1\) against projections based on the ratio
between bins in the background data and the counts in bin \(i\); the deviations
of the projections are given in the SCR vector \(\mathbf{s}\).  Since
\(\mathbf{s}\) is computed using ratios between bins, it is insensitive to
global changes in count rate unless they alter the spectral shape.  This
important feature allows for comparisons of spectra made with unequal
observation times, which happens when numerous observations over wide areas are
aggregated into spatial cells.

\subsubsection{Characterizing Expected Variation}\label{expected-var}

To accurately detect anomalies, we first need to determine what variation can be
expected in the natural background. First, we assume that counts in each energy
bin follow an overdispersed Poisson distribution. As counts are aggregated
across larger spatial cells, they become overdispersed; see
Fig.~\ref{poisson-dispersion}. Consequently, we approximate that \(\text{var}
(c_i) = V c_i\), where \(V\) is the average variance-to-mean ratio of count
rates (\(V=1\) for perfectly Poisson-distributed counts). \(V\) is selected
empirically by calculating the ratio for the area of interest. We know the
relationship between the variance of random variables:
\begin{multline}
\text{var}(aX + bY) = \\
a^2 \text{var}(X)  + b^2 \text{var}(Y) +  2 ab\, \text{cov}(X,Y)
\end{multline}
Hence, using (\ref{scr}) we may estimate \(\text{var}(s_i)\):
\begin{equation}
  \text{var}(s_i) = V c_1 + \left(\frac{b_1}{b_i}\right)^2 V c_i - 2
  \frac{b_1}{b_i} \text{cov}(c_1, c_i)
\end{equation}
Here \(\mathbf{b}\) is treated as an exact value rather than having its variance
propagated. This is a significant simplification, as (\ref{scr}) is nonlinear in
\(\mathbf{b}\) and an exact variance estimate could not be
derived. Consequently, \(\text{var}(s_i)\) will be underestimated. The impact of
this will be explored in Section~\ref{analysis}.

The covariance \(\text{cov}(c_1, c_i)\) would be best estimated from all
previous background observations made in a spatial cell, so that
\(\text{cov}(c_1, c_i) \propto \text{cov}(b_1, b_i)\). However, spatial mapping
would be impractical if it required numerous repeated observations before
detecting anomalies. The relationship between the covariance \(\text{cov}(b_i,
b_j)\) and the correlation \(\text{corr}(b_i, b_j)\) is \cite{Morrison}:

\begin{equation}
  \text{cov}(b_i, b_j) = \text{corr}(b_i, b_j) \sqrt{\text{var}(b_i) \, \text{var}(b_j)}
\end{equation}

To replace \(\text{cov}(c_1, c_i)\) we must also rescale; \(\mathbf{b}\) may
have resulted from an observation of a different duration than
\(\mathbf{c}\). Consequently, we can replace the covariance with a correlation:
\begin{multline}\label{vars}
  \text{var}(s_i) = \\
  V\left(c_1 + \left(\frac{b_1}{b_i}\right)^2 c_i - 2 \frac{b_1}{b_i}
  \left(\frac{T_c}{T_b}\right)^2 \text{corr}(b_1, b_i) \sqrt{b_1 b_i}\right)
\end{multline}
where \(T_c\) is the time taken to observe \(\mathbf{c}\) and \(T_b\) the time
taken to observe \(\mathbf{b}\). This is obtained by replacing
\(\text{var}(b_i)\) with \(\text{var}(b_i T_c/T_b)\), and likewise for
\(\text{var}(b_j)\), rescaling the observation to match the new observation
time.

To compute \(\text{corr}(b_1,b_i)\) we do not rely only on background
observations made in the chosen spatial cell, as there is frequently not enough
data to make this possible. Instead, \(\text{corr}(b_1,b_i)\) is estimated from
observations in all spatial cells by summing together observations into
thirty-second intervals. The correlation between energy bins in these summed
observations is easily computed.

Finally, we can construct the covariance matrix \(\Sigma\) between energy bins
in the SCR vector:
\begin{equation}
\Sigma_{ij} = \text{corr} (s_i, s_j) \sqrt{\text{var}(s_i) \text{var}(s_j)} 
\end{equation}
where \(\text{corr} (s_i, s_j)\) is estimated by summing all observations across
all spatial cells in the same way as for \(\text{corr}(b_i,b_1)\), then
comparing each observation to the global mean spectrum to produce
\(\mathbf{s}\).

By using assumptions about the distribution of the data, we have avoided
requiring direct calculation of \(\Sigma_{ij}\) from the data, as required by
past SCR implementations, which would be impossible when there are too few
background observations. To produce well-conditioned and invertible covariance
matrices, many more observations than variables are required. This would require
numerous repeated mapping passes before any anomalies can be detected, making
the system impractical. Approaches based on shrinkage estimators are possible
\cite{Daniels:2001et}, but cannot work if there are only one or two
observations.

The correlation matrices may be computed and reused for anomaly detection in all
spatial cells; in very large regions where spectra vary greatly, the correlation
matrices may be estimated separately for smaller areas within the monitored
region. Any similar method using covariance matrices would require each spatial
cell to contain observations of equal length, or require new covariance matrices
to be computed for each spatial cell.

\subsubsection{Anomaly Detection}

Previous work has developed a simple anomaly detection algorithm which makes use
of the SCR vector \(\mathbf{s}\) \cite{Pfund:2007}. In these applications a set
of many independent background observations \(\mathbf{b}\) are made, and
\(\mathbf{s}\) is calculated for each observation. After computing the
covariance matrix \(\Sigma\) for the resulting set of SCR vectors,
\(\mathbf{s}\) is calculated for the new observation and it is compared to the
background through the mathematical construct of the Mahalanobis distance:

\begin{equation}
\label{dist}
D^2 = \mathbf{s}^T \Sigma^{-1} \mathbf{s}
\end{equation}

The Mahalanobis distance measures the difference between a multivariate
observation and the typical mean, normalizing by the typical variance expressed
in the covariance matrix \cite{Morrison}. This implies that spectral shape
changes consistent with already-observed natural background variations will
cause only small increases in \(D^2\), while changes very different from the
past data will produce large discrepancies.

If the estimated covariance matrix \(\Sigma\) is accurate and the background
radiation source is unchanging, the Mahalanobis distance \(D^2\) should be
\(\chi^2\)-distributed with \((n-1\)) degrees of freedom \cite{Pfund:2010hm}. In
practice there are slight background fluctuations from various natural
processes, and the distribution departs slightly from theoretical predictions
\cite{Pfund:2010hm}. (See Fig.~\ref{scr-distribution}.)

By setting a desired alarm threshold \(D_A\) based on typical natural spectral
variations, one can search for unusually large spectral anomalies, which may
indicate source changes.  To monitor a wide area, observations can be aggregated
into spatial cells and the cumulative spectrum in each cell compared to previous
observations. Each cell may contain different numbers of observations, and an
individual cell may have observations at different times and locations from day
to day.

Due to (\ref{vars}), as \(T_c\) increases \(\text{var}(s_i)\) decreases. This
renders the system more and more sensitive with longer observation times.

The method does not distinguish between equally-sized anomalies from natural or
artificial sources, though the choice of energy bins may be made to optimize
sensitivity to certain isotopes \cite{Pfund:2007}. Additional benefits of this
approach include being able to detect unexpected decreases in radioactivity as
well as the replacement of one radioactive source with another, which may
maintain a constant count rate in a given area while altering the spectral
shape.

\section{Example Dataset}
We collected a sample dataset between June 22nd and August 10th, 2012 at the
University of Texas's J.~J.~Pickle Research Campus.  To survey the campus, the
system was placed in a golf cart and driven on most work days on irregular
routes. Total observation time amounted to twenty hours in 48 observation runs
containing a total of more than 37,000 individual two-second observations.

\begin{figure}
  \centering
  \includegraphics[width=3.5in]{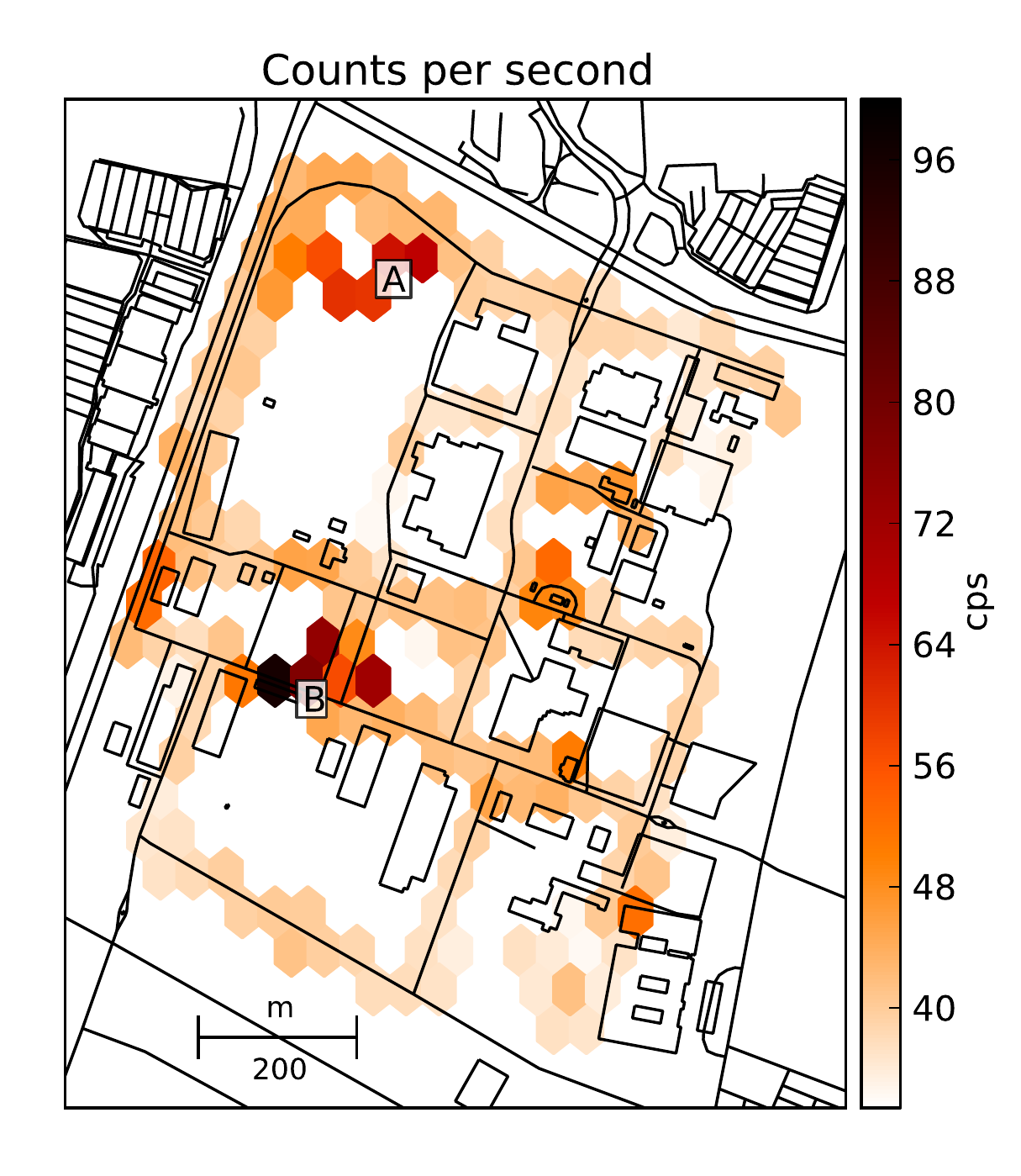}
  \caption{A map of J.~J.~Pickle Research Campus with gamma counts per second
    overlaid, averaged over one month of mapping. Areas of elevated background
    include the radioactive materials storage facility at the northwest corner
    (A) and a cluster of large brick buildings near central campus (B).}
  \label{prc-cps}
\end{figure}

The natural gamma background varies spatially and temporally due to many natural
factors.  Our surveys revealed a spatially varying natural background across
Pickle Research Campus. A map of average radiation levels over several weeks is
shown in Fig.\ \ref{prc-cps}. Background activities vary spatially by a factor
of two due to natural and artificial sources.

\section{Analysis}
\label{analysis}

\subsection{Poisson Distribution Assumption}

Our algorithmic approach relies on the underlying Poisson distribution of
radiation data.  To test the validity of this assumption, Poisson dispersion
tests were performed on the dataset as a function of spatial scales. The Poisson
dispersion test determines whether a given set of observations could plausibly
have been drawn from the same Poisson distribution \cite{Rao:1956vp}. The test
computes a dispersion parameter \(P\), defined by:
\begin{equation}
P = \frac{1}{\bar x} \sum_{i=1}^N (x_i - \bar{x})^2
\end{equation}
where \(\bar{x}\) is the mean of all observations \(x_i\) and \(N\) the total
number of observations. This parameter is \(\chi^2\)-distributed with \((N -
1)\) degrees of freedom. Using this distribution, a \(p\) value can be computed
for \(P\), indicating the probability that the observed distribution of values
would arise from a Poisson-distributed random variable. For example, dividing
our dataset into 125-meter grid cells, we find that on average, \(p = 0.33\), so
we do not reject the hypothesis that the data are Poisson distributed; this
falls to \(p = 0.17\) for 250-meter cells.

\begin{figure}
  \centering
  \includegraphics[width=3.5in]{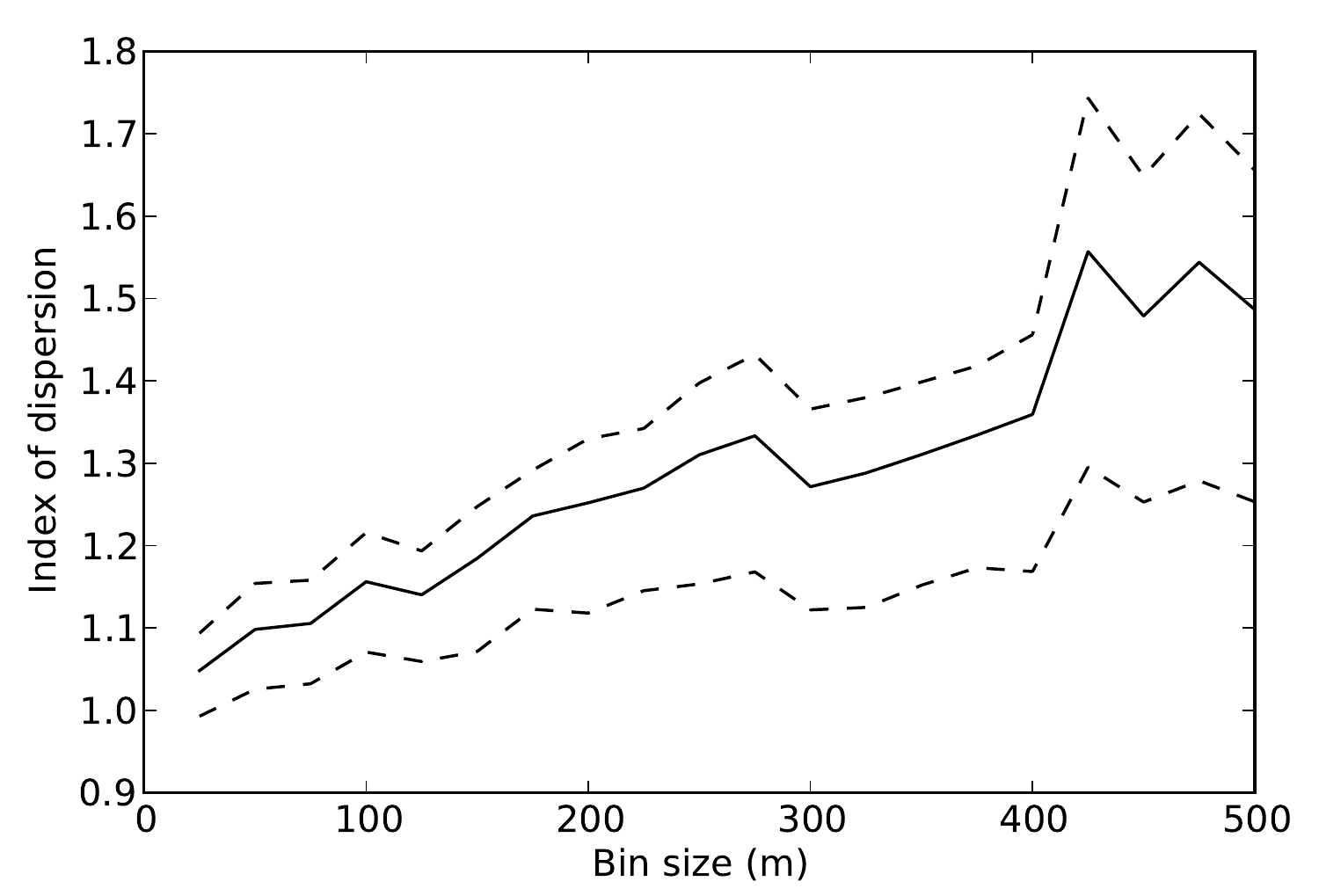}
  \caption{Mean, minimum and maximum index of dispersion of counts in each
    energy bin for all grid cells, at specified grid sizes. An index of 1 is
    consistent with a Poisson distribution.  As expected, smaller cells more
    closely match the Poisson distribution, while larger cells have additional
    variance.}
  \label{poisson-dispersion}
\end{figure}

To quantify this, the index of dispersion was computed for all grid cells at
various cell sizes. The index of dispersion \(V\) is a measure of the variance
of a distribution, compared to its mean:
\begin{equation}
V = \frac{\sigma^2}{\mu}
\end{equation}
where \(\mu\) is the distribution's mean and \(\sigma^2\) its variance. The
variance of a Poisson distribution equals its mean, so the index of dispersion
is expected to be one. Fig.\ \ref{poisson-dispersion} shows mean indices of
dispersion for various grid cell sizes, demonstrating that smaller cells tend to
have count rate distributions closer to the expected Poisson distribution, as
expected. To account for this, we adjusted the dispersion parameter \(V\) in
Eq.~\ref{vars} to match the mean index of dispersion for grid cells at a chosen
spatial scale.

Of course, this problem is dependent on the nature of the area being mapped;
areas with greater spectral variation on short spatial distances will need to be
mapped with smaller spatial cells. 

\subsection{Comparing Spatial and Temporal Variation}

\begin{figure}
  \centering
  \includegraphics[width=3.5in]{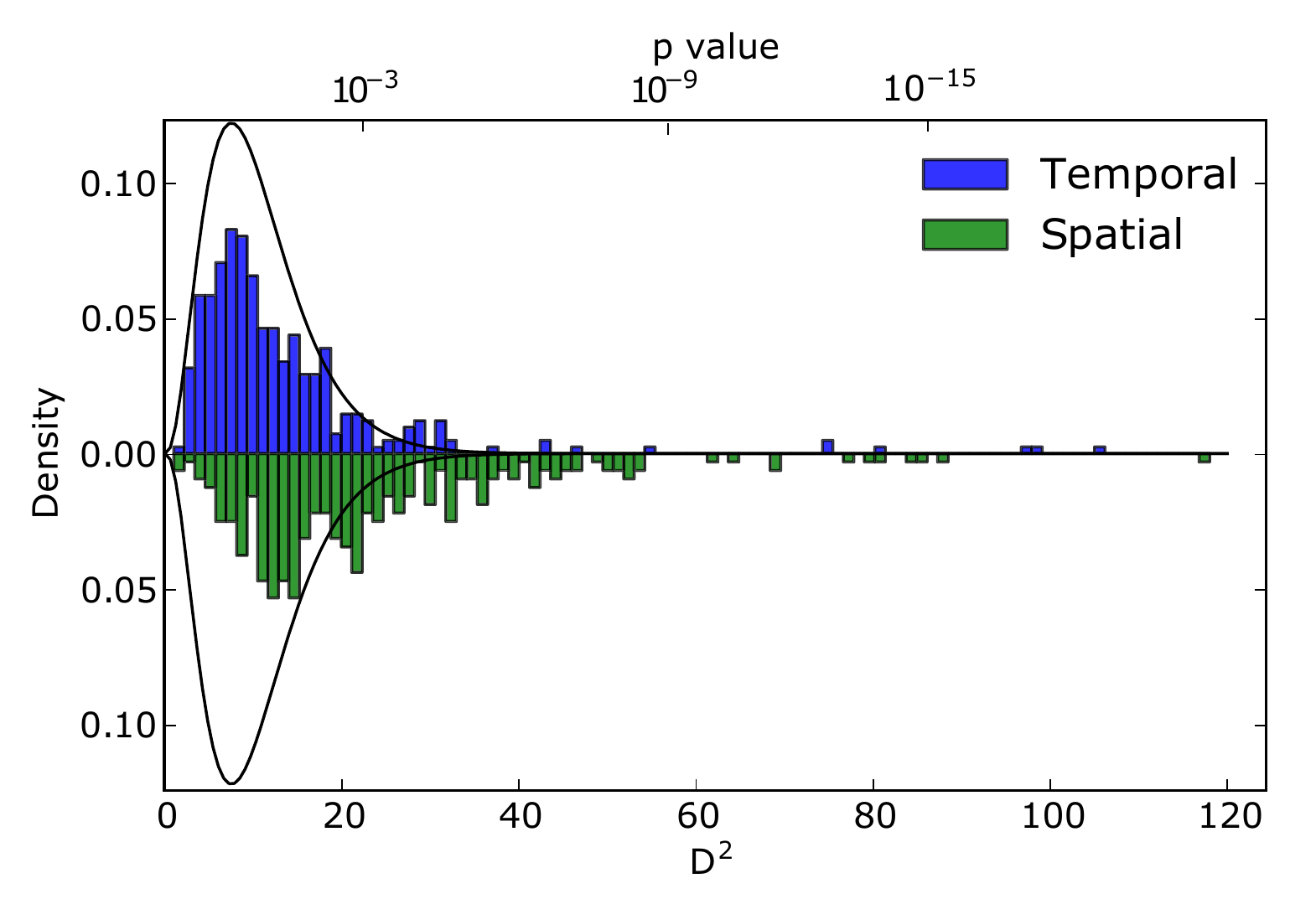}
  \caption{Distribution of the SCR anomaly statistic \(D^2\) in 250-meter grid
    cells in the sample dataset, with a \(\chi^2\) distribution (with 7 degrees
    of freedom) overlaid. The upper histogram is the result of comparing each
    grid cell to the same cell on the previous day; the lower histogram compares
    each cell to a fixed reference cell, showing spatial rather than temporal
    variation. The lower distribution is clearly shifted to the right.}
  \label{scr-distribution}
\end{figure}

Our dataset at the Pickle Research Campus revealed not only spatial but temporal
variation in background. To compare the temporal variation to the spatial
variation, the observation area was divided into grid cells 250 meters on each
side, and each day's set of observations was compared to two or more previous
days using the SCRAM algorithm. Fig.\ \ref{scr-distribution} shows the
distribution of \(D^2\) recorded over several dozen passes through the area
during July 2012. For comparison, it also plots the \(D^2\) values calculated by
comparing each grid cell's spectrum to a fixed reference cell, rather than
comparing it to the same cell on a previous day.

The spatial distribution is clearly shifted to the right, indicating that there
is more spatial variance than temporal variance. This validates the assumption
of the SCRAM method that it is best to compare spectral observations to previous
observations made in the same place, rather than observations made at other
locations. In other words, the background spectra vary much more spatially than
they do temporally. This suggests that alarm thresholds may be set lower when
using the SCRAM method to perform anomaly mapping, giving higher sensitivity
without increased false alarm rates; knowledge of the prior background is more
useful than the detector's knowledge of what it observed thirty seconds ago at a
different location, allowing more sensitive detection.

Further research should also investigate the impact of known natural variations
in background radiation, such as those caused by rainfall's interaction with
radon daughter products, which have been extensively modeled
\cite{Mercier:2009gv}. At least five inches of rainfall occurred during our
observation period, so these effects may be present in our example dataset,
accounting for some of the temporal variance, and can be compared against
existing models.

The presence of \(p < 10^{-3}\) anomalies in Fig.~\ref{scr-distribution} is not
unusual. The \(p\)-values are computed under the \(\chi^2\) distribution, which
the data do not follow perfectly: There are, of course, some true variations in
natural background in the data, and the distribution of counts may be more or
less overdispersed in different spatial cells. As noted in
Section~\ref{expected-var}, estimates of \(\text{var}(s_i)\) are systematically
low, making some observations appear more anomalous than they should. Also, some
spectral comparisons were made with little data -- because of irregular routes,
a spatial cell may only contain a few seconds of data, giving an extremely poor
estimate of the spectrum in that cell. Regardless of these issues, the algorithm
is practical for anomaly detection, as demonstrated in the following
simulations.

\subsection{Detection Performance}

\begin{figure}
  \centering
  \includegraphics[width=3.5in]{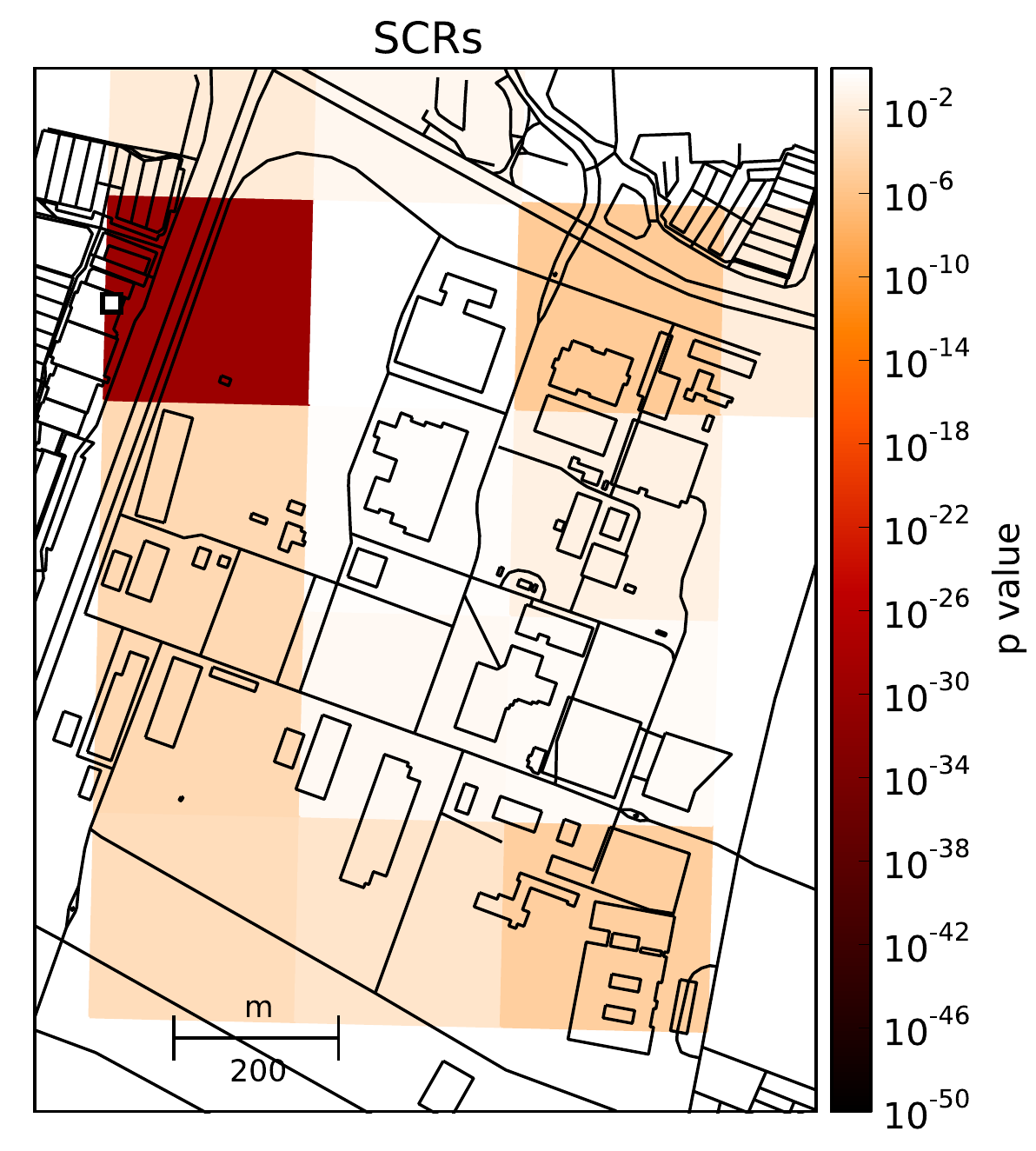}
  \caption{A heatmap of SCR values comparing one day to the previous days, with
    $p$ values computed based on the \(\chi^2\) distribution of \(D^2\). A
    simulated 140 mCi cesium-137 source was injected on the west side of campus,
    at the small square. The detector traveled along the adjacent road, never
    closer than 100 meters.}
  \label{scr-injected-map}
\end{figure}

To demonstrate the expected performance of the SCRAM algorithm, we performed
several simulations. Simulations used real observed spectra from a \(0.844 \,
\mu\text{Ci}\) cesium-137 check source. The simulation code was calibrated by
placing the source at known distances from the scintillator detector, and
accounted for geometric attenuation of count rates, as well as the exponential
attenuation due to gamma absorption in air.

The first simulation, shown in Fig.~\ref{scr-injected-map}, injected a simulated
140 mCi cesium-137 source on the west side of Pickle Research Campus, 100 meters
from the detector's typical path. The detector, driving past at roughly 10 miles
per hour, easily detected the source with \(D^2 = 156\) (a far larger anomaly
than any natural variation recorded in Fig.~\ref{scr-distribution}),
demonstrating the ability of a small moving detector to detect sources at
distance.

Next, we performed a simulation to calculate the minimum detectable radioactive
source size at a variety of distances. We selected the same straight stretch of
road on the northwest corner of the Pickle Research Campus and simulated the
injection of a cesium-137 source at various distances from the road into the
collected background data.

\begin{figure}
  \centering
  \includegraphics[width=3.5in]{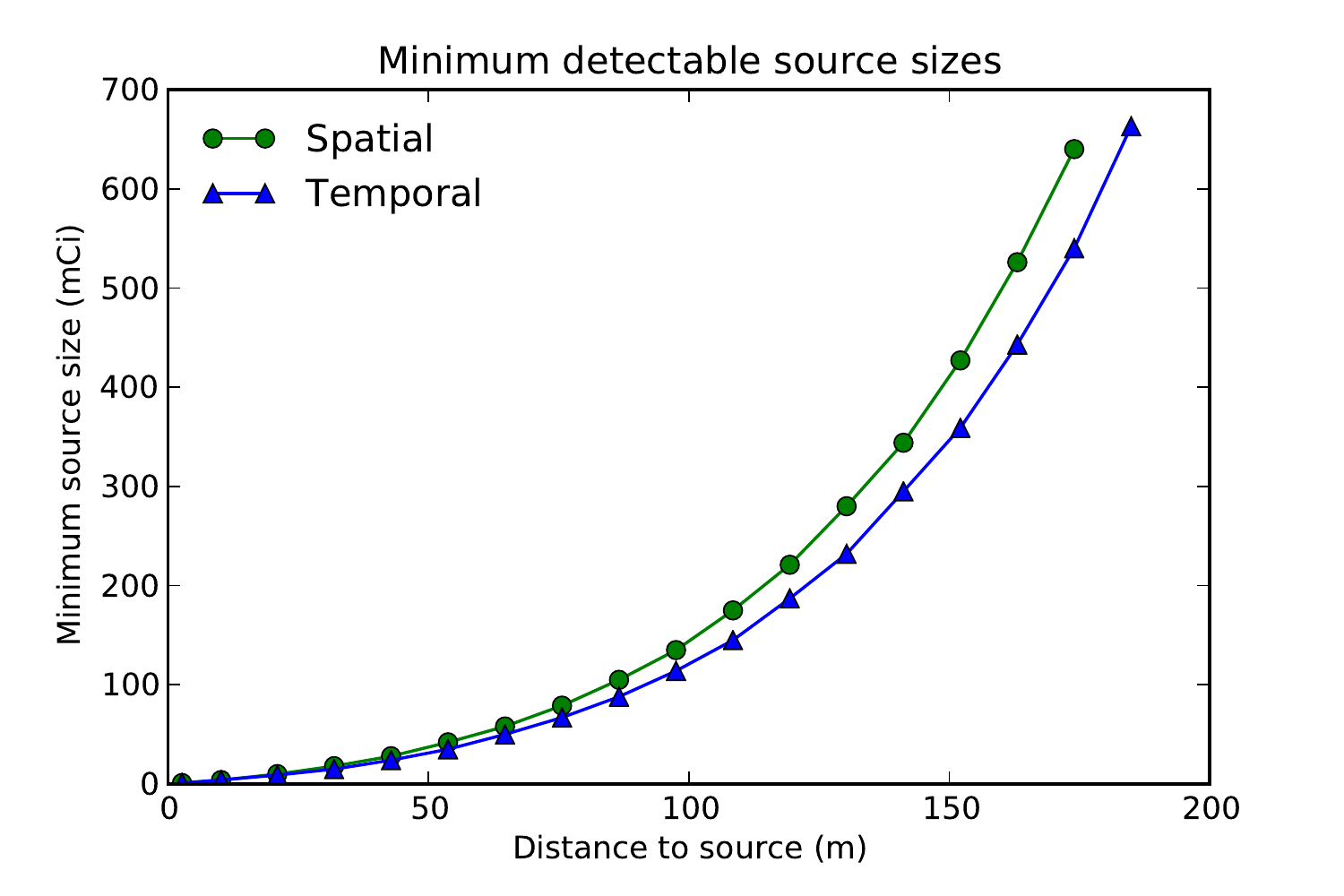}
  \caption{The minimum cesium-137 source sizes required for detection at least
    80\% of the time, at various distances away from the detector's path, using
    both temporal and spatial comparisons of spectra. Total observation time was
    136 seconds.}
  \label{minimum-detectable}
\end{figure}

To choose our alarm threshold, we used the anomaly statistics distribution data
shown in Fig.~\ref{scr-distribution}, selecting a threshold which only 1\% of
our example dataset exceeds. This corresponds to a 1\% false alarm rate. For
temporal comparisons of spectra, this threshold is \(D_A^2 = 83\), while for
purely spatial comparisons the result is \(D_A^2 = 113\). We computed the
minimum source size required to achieve a statistical power of 0.8 (that is,
detection of a true positive at least 80\% of the time) for both thresholds, and
the results are given in Fig.~\ref{minimum-detectable}.

These results show that taking advantage of the exact prior background spectrum
at each location leads to approximately 20\% better detection performance over
the use of spectra from a single fixed reference point. Of course, smaller
sources would be detectable with a much larger scintillator or with a much
longer observation time.

\subsection{New and Improved?}

There are many academic, prototype, and commercial systems that map
radiation. The SCRAM system contains many elements in common with previous
demonstrations, including the detector, GPS, spectral binning, spectral
comparison algorithm, and spatial database. For these components, we are
agnostic to replacements: different detectors or spectral binning choices could
be chosen for specific applications. The novel component of the SCRAM system is
estimation of the local background and its variance from previous observations.
As has been pointed out previously, simply deploying larger detectors does not
improve detection performance, as the local background is the limiting factor
\cite{Ziock:2002bo}.

It is instructive to compare the SCRAM system against commercial systems and
federal guidelines. However, these comparisons are difficult, given varying
detector sizes and modes of operation. The Domestic Nuclear Detection Office has
published testing methodologies we plan to follow as our system evolves from a
proof-of-concept \cite{DNDO}. These guidelines require less than one false alarm
per hour for 60-second integrations, amounting to a 1.7\% false alarm rate. We
chose a more strict limit of 1\%, though this is not directly comparable; our
spectral comparisons are made per spatial cell, rather than per 60-second
integration. This limit can be chosen to meet user requirements.

For detection performance comparison, we can look to mapping systems with
similar goals. The commercial ORTEC NaI-SS consists of a \(4\times 4 \times
16^{\prime\prime}\) sodium iodide detector transported by car, recording spatial
and spectral data every second \cite{ORTEC}. For ORTEC's chosen false alarm
rate, the detector can identify a 10 \(\mu\)Ci cesium-137 source at 3 meters
while traveling at 10 miles per hour. Similarly, the SORIS coded aperture imager
uses four large van-mounted sodium iodide plate scintillators, and can detect 1
mCi cesium-137 sources at a distance of 100 meters
\cite{Zelakiewicz:2011ig}. The detector in the SCRAM system is hundreds of times
smaller; a simple scaling of our performance by collection area ignores many
factors, but nevertheless, the SCRAM concept shows its relevance and promise by
displaying significant gains over these previous detection limits when scaled to
their detector sizes. Given all the differences in the systems, direct
quantitative comparisons are not currently in order.

\section{Conclusion}

In this work, we report on the development of an integrated system for wide-area
radiation surveillance. A mobile detector is used to collect multiple-pass data,
which is stored into a spatial-temporal spectral database. We developed a novel
approach for anomaly detection of the spectral content by comparing observations
to previous determinations of the background. Because the spatial variance is
much larger than the temporal variance, this multi-pass methodology has the
potential to deliver increased sensitivity to weak or distant sources, as
demonstrated by simulations.

These developments are the first steps necessary to implement the larger vision
of a providing continuous wide-area surveillance through the use of mobile
sensors. Future improvements to the algorithm include the use of spatial kriging
to improve estimates of background spectral shape, methods to determine optimal
energy bin sizes, and adaptation for use in small, low-power mobile
detectors. 

\section*{Acknowledgment}

The authors would like to thank T.~Tipping of the Nuclear Engineering Teaching
Laboratory and S.~Pennington of Environmental Health and Safety at the
University of Texas for their valuable suggestions and support, J.~D.~Biniaz and
P.~Vetter for their suggestions on the context and need for wide-area radiation
surveillance, and staff at Bridgeport Instruments for technical support and
advice regarding use of their detectors. The authors would also like to thank
R.~Schwitters for the suggestion to plot anomalies as \(p\) values, as well as
several other helpful suggestions.

\bibliographystyle{ieeetr}
\bibliography{algorithms}

\end{document}